# Experimental realization of wide-mode-area slow light modes in valley photonic crystal heterostructure waveguides

CHENGKUN ZHANG,[1,2,*] GUANGTAI LU,[1,2] NATTHAJUKS PHOLSEN,[1,2] YASUTOMO OTA,[3] AND SATOSHI IWAMOTO [1,2]

[1]*Research Center for Advanced Science and Technology, The University of Tokyo, 4-6-1 Komaba, Meguro-ku, Tokyo 153-8904, Japan*
[2]*Institute of Industrial Science, The University of Tokyo, 4-6-1 Komaba, Meguro-ku, Tokyo 153-8505, Japan*
[3]*Department of Applied Physics and Physico-Informatics, Faculty of Science and Technology, Keio University, 3-14-1 Hiyoshi, Kohoku-ku, Yokohama-shi, Kanagawa 223-8522, Japan*
*\*zck20@iis.u-tokyo.ac.jp*

**Abstract:** We experimentally realized wide-mode-area slow-light modes in valley photonic crystals (VPhCs) heterostructure waveguides. The waveguides are fabricated on a silicon slab by inserting gapless photonic graphene layers with varying widths and modifying the unit cell spacing near the domain walls. By reducing the spacing between unit cells at the domain boundaries, slow-light guided modes are achieved in VPhCs heterostructure waveguides. The presence of wide-mode-area modes is verified by observing the radiation in light propagation of leaky guided modes above the light line. To characterize guided modes below the light line, we introduce air-slot terminations to induce out-of-plane scattering and measure intensity profiles. The results show that the mode widths are tunable for both fast-light and slow-light modes in VPhCs heterostructure waveguides by adjusting the number of photonic graphene layers. The ability to support wide-mode-area slow-light modes in VPhC heterostructures offers promising opportunities for the development of high-power, on-chip photonic integrated devices.

## 1. Introduction

Topological photonics is a rapidly emerging field that applies the concepts of topological insulators to engineered optical structures, exploring novel ways to control the flow of light [1-10]. A hallmark of this field is the existence of topologically protected edge state-electromagnetic modes localized at the interfaces between photonic media of distinct topological phases, which propagate unidirectionally along domain walls and are remarkably robust against scattering from defects or disorder, as ensured by the nontrivial bulk band topology of the surrounding structures [11-21]. In two-dimensional photonic systems, a variety of platforms have been explored to realize such protected light transport [11-21]. For instance, magneto-optical photonic crystals (PhCs) under an external field can emulate the quantum Hall effects of light, exhibiting one-way edge propagation without backscattering [11, 14, 15, 18]. Alternatively, all-dielectric photonic lattices that break certain spatial symmetries have achieved analogues of quantum spin Hall and quantum valley Hall phases, where pairs of counter-propagating edge modes carry opposite pseudospins or reside in different momentum valleys [12, 13, 16, 17, 19-23]. Within this landscape, VPhCs have emerged as a particularly promising realization of topological edge transport due to their compact sizes and low-radiation-loss transmission [21-27]. Utilizing the kink edge states in VPhCs, numerous photonic devices including optical waveguides [24-27], optical power splitter [28-31], and lasers [32-34] have been demonstrated, providing fruitful components for photonic integrated circuits.

Generally, the topological edge states including valley kink edge states are tightly localized near the interface between two PhCs with distinct topology [11-13, 18-23]. Even though these localized edge states have some advantages like enhancing nonlinear effect, the narrow mode width also limits the power transmission in waveguides [35, 36], which hinders on-chip high-power devices. To address this issue, recently a PhC heterostructure supporting topological chiral edge states with a large mode width has been realized in magnetic PhCs, by inserting an unmagnetized PhCs in center region [35]. The topological one-way large-area waveguide can transport electromagnetic field energy efficiently and provide a novel route to control electromagnetic waves. A similar concept has been realized in VPhC platforms by inserting gapless photonic graphene layers between two VPhCs with opposite topology, leading to the formation of the valley photonic crystal heterostructure waveguides (VPHW) [36]. The proposed VPHW allows tunable mode widths of topological single guided modes by varying the size of the inserted photonic graphene region, facilitating high-power transmission with wide-mode-area single-mode operation [36]. By using the tunable mode widths in VPHWs, various topological photonic devices have been realized or proposed, including beam splitters with controllable splitting ratios [37-40], energy concentrators and taper-free waveguides based on abrupt mode-width transitions [41-43], as well as multiplexers for optical communication [44, 45]. Notably, all these devices rely on guided modes with low group indices, while the realization of slow-light modes with rich functionalities in VPHWs remains unexplored.

Slow-light waveguides in PhCs offer a promising platform for compact optical delay lines, high-speed modulators, and efficient nonlinear optical devices, owing to their ability to reduce the group velocity of light and enhance light–matter interactions [46-48]. Slow-light modes have been extensively studied in W1 PhCs waveguides for on-chip delay lines and modulators [49, 50]. In addition, slow-light waveguides based on topological VPhCs with bearded interfaces have been demonstrated, exhibiting robustness against sharp bends [26, 27]. A common feature for conventional and topological slow-light waveguides discussed above is that the slow-light modes are tightly localized near interface, and wide-mode-area slow-light modes are absent, which are attractive because of their capacity of high-power transmission while keeping slow-light properties. To combine slow-light modes and large mode width, previously, we have reported a way to realize wide-mode-area slow-light single modes in an air-in-slab VPhC heterostructure [51]. We found that the group velocity of the topological guided mode can be significantly reduced by modifying the spacing between PhCs unit cells near the domain walls, thereby realizing slow-light single guided modes. Additionally, the mode width of the slow-light guided mode in the VPhC heterostructure can be controlled by adjusting the width of the gapless photonic graphene layers.

In this paper, for the first time, we experimentally realized wide-mode-area slow-light modes in VPhC heterostructure waveguides. We fabricated VPhC heterostructure waveguides in a silicon slab by inserting gapless photonic graphene layers of varying widths and adjusting the unit cell spacing near the domain walls. Transmission spectra were measured for the fabricated samples, and group indices exceeding 20 were extracted, confirming the existence of slow-light modes. Additionally, optical imaging shows that the leaky guided modes broaden as more photonic graphene layers are inserted. To probe guided modes below the light line, the VPhC heterostructure waveguides are terminated to induce scattering, enabling the measurement of intensity profiles. These measurements demonstrate that the mode widths of both fast-light and slow-light guided modes can be effectively tuned by varying the number of photonic graphene layers at telecommunication wavelength. This study establishes a new foundation for integrating slow-light functionalities with large mode areas into high-power photonic integrated circuits.

## 2. Results and discussion

### 2.1 VPhC heterostructure: structure design and sample fabrication

Our designed structure is based on a VPhC heterostructure $A|B_N|C$, consisting of three domains: a graphene-like PhC B domain with $N$ layers sandwiched by two domains of VPhC A and VPhC C, as illustrated in Fig. 1(a). The domain walls between the graphene-like PhC B and VPhC A, C form zigzag-like interfaces. Figure 1(b) depicts the unit cells of three domains, where each unit cell is a silicon structure ($\varepsilon = 11.56$) with six triangular air holes ($\varepsilon = 1$) forming a hexagonal lattice. The side lengths of large and small air holes in VPhC A and C are $L_L = 1.3a/\sqrt{3}$ and $L_S = 0.7a/\sqrt{3}$, while the side length of all air holes in PhC B is $L_L = L_S = 0.95a/\sqrt{3}$. ($a$ is period of PhC). To achieve slow-light modes, the value $d$, distance between PhC unit cells near the domain walls, is reduced. As an example, Fig. 1(c) depicts top view of VPhC heterostructure $N = 4$ with $d = 1.4a$ and $d = 1.2a$. It is noticed that the silicon slab is compact even if $d$ is reduced, which keeps possibility for fabrication. In Fig. 1(d) (e), the projected band structures of transverse electric (TE) mode for two structures in Fig. 1(c) are plotted, where single guided modes region is marked by red. The calculation is in three-dimensional (3D) plane-wave expansion (PWE) method for $a = 550$ nm and 220-nm-thick silicon slab. When $d = 1.4a$, a topological guided mode with linear dispersion, along with several trivial guided modes, appears within the bandgap of the VPhCs. As $d$ is reduced to $1.2a$, the guided modes near the K point shift to higher frequencies due to the local refractive index reduction near the domain walls, while the modes near the Brillouin zone center remain largely unaffected, as they are primarily confined to the central region of the VPhC heterostructure [51]. This imbalanced change results in the emergence of slow-light guided modes in VPHW as shown in Fig. 1(e). Figure 1(f) shows the amplitude distribution of the magnetic field component perpendicular to the PhC plane ($H_z$) for a guided mode with $n_g \sim 50$, indicated by a red point in Fig. 1(e). The field is observed to extend across the entire domain of PhC B and decay into the adjacent VPhC A and C domains, resulting in a broader mode profile compared to slow-light modes in interface-based waveguides. We emphasize that reducing $d$ can lead to the emergence of slow-light modes across different values of $N$ [51]. Moreover, the mode field of the slow-light modes can be further broadened by increasing $N$, thereby enabling the realization of wide-mode-area slow-light guided modes in the VPHW.

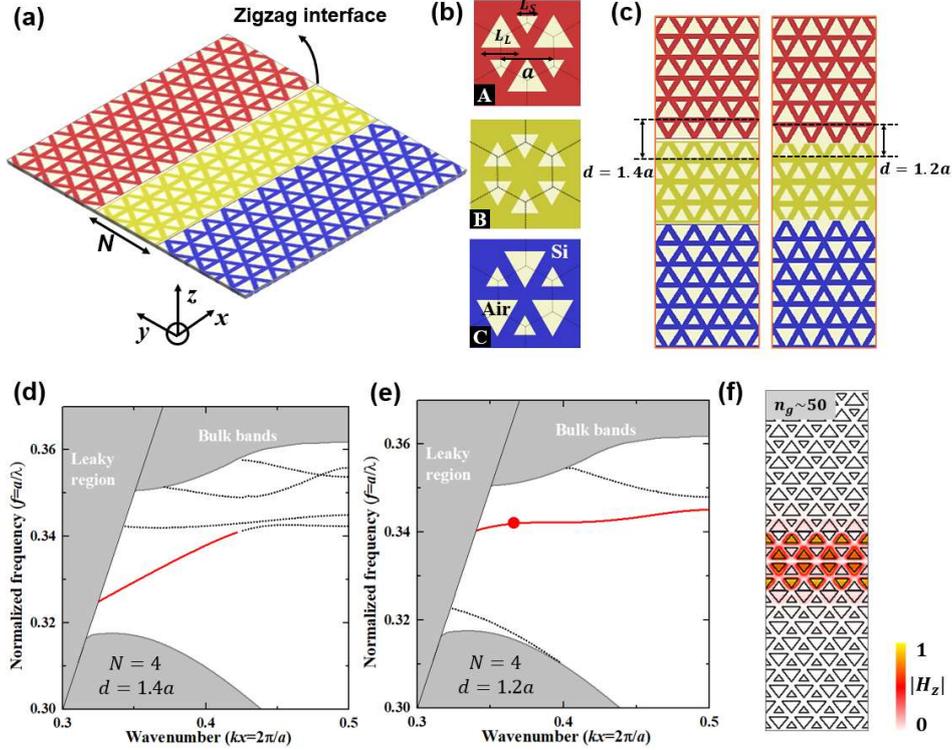

Fig. 1. (a) Schematic illustration of the VPhC heterostructure A|B$_N$|C consisting of three domains, VPhC A, PhC B, and VPhC C. This figure shows a structure with $N = 4$ as an example. (b) Schematic diagrams for unit cells of VPhC A, PhC B, and VPhC C. (c) Top view of VPhC heterostructure $N = 4$ with $d = 1.4a$ and $d = 1.2a$. (d) (e) Projected band structures for designed structures in (c). The calculation is in 3D PWE method. (f) Amplitude distribution of mode width $n_g \sim 50$ marked as red point in (e).

We fabricated the designed VPhC heterostructure waveguides in silicon-on-insulator platform using electron beam lithography and reactive ion etching. Air-bridged structures were realized by removing the sacrificial silicon dioxide layer beneath the silicon slab through vapor etching. A lattice constant of 550 nm was chosen to align the operating wavelength with the telecommunication band. In experiment, we fabricated VPhC heterostructure waveguides with different $N$ and $d$ values for characterizing the optical properties. As an example, Fig. 2(a) and (b) depict scanning electron microscopy (SEM) images of fabricated VPhC heterostructure waveguides for $N = 4, d = 1.4a$ and $N = 4, d = 1.2a$, where the VPhC A, PhC B, VPhC C domains are highlighted in red, yellow, and blue respectively. The inset in Fig. 2(b) shows an enlarged view of the structure near the domain wall, where two PhC unit cells are connected and local refractive index is reduced. Two semicircular grating couplers are placed to terminate the waveguides for light input and output. The grating couplers are designed to enhance reflectance at the waveguide end, leading to clear Fabry-Perot (FP) fringes in transmission spectra for group index calculation [52, 53].

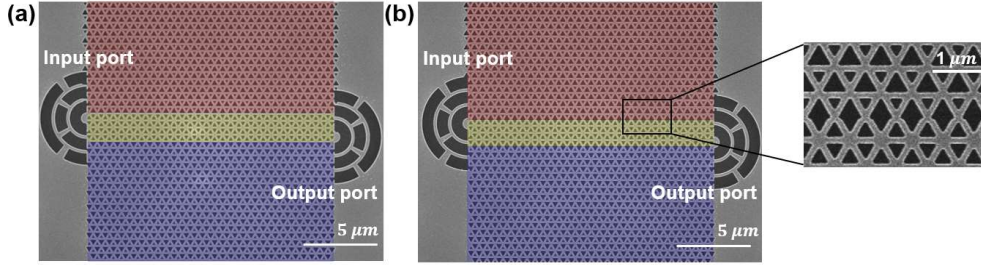

Fig. 2. SEM images of fabricated VPhC heterostructure waveguides with (a) $N = 4, d = 1.4a$ (b) $N = 4, d = 1.2a$. The inset in (b) shows the structure near the domain walls between VPhC and PhC B.

## 2.2 Optical transmission measurements

To characterize the optical properties of guided modes, we measured the transmission spectra and extracted group indices for VPhC heterostructure waveguides with different $N$ and $d$. In the measurement, a laser-driven xenon lamp covering wavelengths of guided modes was used. The input light passes through a 50X objective lens ($\text{N.A.} = 0.65$) from the top of the sample and focuses on the input grating port. Output light from the other grating coupler is collected from the objective lens and sent to a spectrometer with an InGaAs cooling camera for spectral analysis. Figure 3(a) shows measured transmission spectrum for VPhC heterostructure waveguide with $N = 4, d = 1.4a$, corresponding to the designed structure whose band structure is shown in Fig. 1(d). High transmittance is observed around 1550 nm, demonstrating compatibility with optical telecommunication applications. It is noticed that sharp peaks occur in the spectrum, which results from FP resonance by the reflectance at the end of waveguide. Group indices were extracted from the measured FP fringes for 30-period-long waveguides and are presented as solid red dots in Fig. 3(b). The measured group indices remain below 10 with minor fluctuations, indicating fast-light guided modes with near-linear dispersion. In Fig. 3(b), we also plot group indices calculated by 3D PWE method as solid black line with an offset in horizontal axis for a better comparison. The measured group indices show good agreement with simulation results. Figure 3(c) shows the measured transmission spectrum of VPhC heterostructure waveguide with $N = 4$ and a reduced unit cell spacing of $d = 1.2a$, corresponding to the designed structure whose band structure is shown in Fig. 1(e). Compared with Fig. 3(a), the transmission spectrum in Fig. 3(c) shows a narrower guided band at shorter wavelength, which originates from the band structure change at Fig. 1(e). We notice that there are two clear band gaps on the sides of guided modes in Fig. 3(c), corresponding to the upper and lower band gaps in simulated band structure shown in Fig. 1(e), indicating consistence between measurement and simulation. As shown in Fig. 3(d), most values in the group indices extracted from Fig. 3(c) exceed 20, confirming the existence of slow-light modes. We also plot the calculated group indices from simulation as solid black line for comparison, and the measured group indices fit well with simulation results. These results manifest that slow-light modes are realized in VPhC heterostructure waveguides by narrowing the spacing between PhC unit cells near the domain walls.

In our previous simulation, we proposed that slow-light modes can be realized by reducing the distance $d$ for VPhC heterostructure waveguides with varying number of layers $N$. To experimentally validate this, we fabricated a series of samples with $N = 4,6,8$ and $10$, each with different $d$. We extracted the group indices from the measured transmission spectra and found that some group indices for guided modes are above 20, which indicates slow-light modes can be realized in VPhC heterostructure waveguides with varying number of layers $N$ by reducing $d$ values (see Supplement 1).

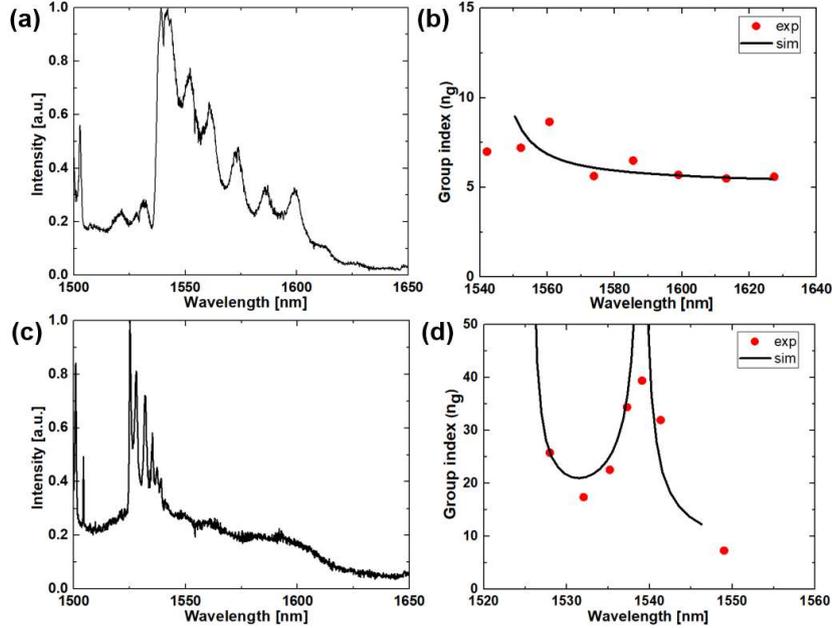

Fig. 3. Measured transmission spectra and group indices for VPhC heterostructure waveguides with $N = 4, d = 1.4a$ for (a) and (c), and $N = 4, d = 1.2a$ for (b) and (d). The black lines in (b) and (d) are calculated group indices in 3D PWE method.

### 2.3 Tunable mode width in light propagation imaging

To experimentally demonstrate wide-mode-area slow-light modes in VPhC heterostructure waveguides, it is crucial to verify not only the existence of slow-light modes but also the tunability of mode width through its increase with $N$. Tunable mode width is an important property of VPhC heterostructure waveguides, which is determined by the number of inserted graphene-like layers, enabling wide-mode-area guided modes. Previous studies have demonstrated that the mode width in VPhC heterostructure waveguides can be controlled by the size of the PhC B region in the microwave regime [36]. However, despite its importance for optical communication, experimental demonstration of tunable mode widths in VPHWs at telecommunication wavelengths remains absent.

Here, we demonstrate tunable mode widths at telecommunication wavelengths by imaging light propagation. In the first experiment, we used an InGaAs camera to image light propagation in VPHW with different $N$, using the light source and samples same as those employed for transmission spectrum measurement. Figure 4 presents near-infrared microscope images of light propagation in VPHW with $N = 2,4,6,8,10,12$ and $d = 1.4a$, corresponding to the designed structures featuring linear guided modes in their band structures. In these images, the large and small spots indicate input and output light, respectively. The samples are indicated by red boxes, and their widths increase with $N$, the number of PhC B layers, while VPhCs remain unchanged. Light propagation between the input and output ports is observed, corresponding to radiation from leaky guided modes propagating above the light line. The radiation becomes broader with increasing $N$, suggesting that the guided mode width in the VPhC heterostructure waveguides increases accordingly. These results intuitively illustrate that tunable mode widths can be achieved for the guided modes in VPhC heterostructure waveguides by varying the number of PhC B layers.

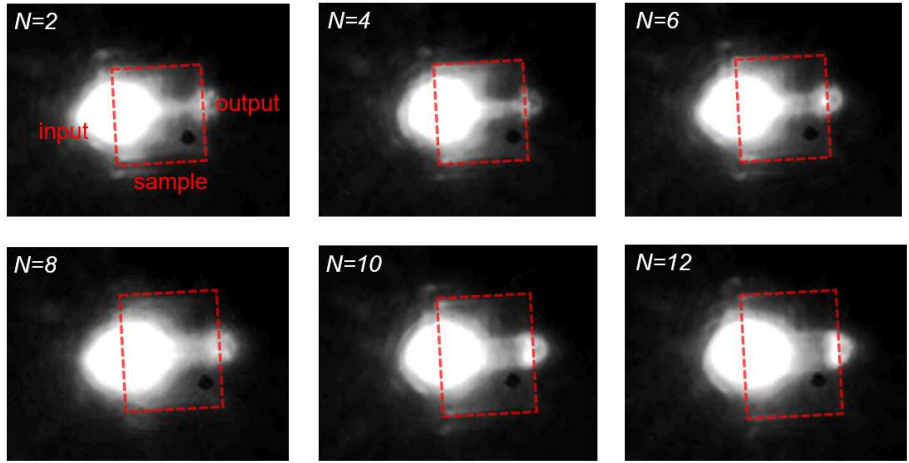

Fig. 4. Near-infrared microscope images of light propagation in VPhC heterostructure waveguides with $N = 2,4,6,8,10,12$ and $d = 1.4a$. The large and small spots correspond to input and output light, and samples are indicated by red boxes.

### 2.4 Mode width measurement by scattering

In the above, we intuitively demonstrated that the mode widths become broader with increasing $N$ in VPHW by imaging light propagation, which results from leaky guided modes above the light line. For real waveguiding devices, guided modes below the light line without radiation are more attractive for significantly reducing the loss. However, these guided modes do not radiate, making direct observation of their propagation challenging. To measure mode widths for guided modes below the light line, we terminated the VPHW with air slots, which introduces strong scattering arising from large refractive index contrast. The samples designed for light scattering are located next to the samples with same $N$ and $d$ employed for spectrum measurement. A wavelength-tunable single-mode laser diode is utilized as a light source and the output power is set at 500nW. By probing intensity profiles of light scattering in VPHW with different $N$, we calculated full-width half-maximum (FWHM) as a measure of mode width, and derived relationship between mode widths and $N$.

Firstly, we measured intensity profiles of light scattering in VPHW with different $N$ and unchanged $d = 1.4a$. Figure 5(a) shows an SEM image of fabricated VPHW with $N = 10$ terminated by an air slot, where a semicircular grating coupler is set for light input. Figure 5(b) shows a microscope image of light propagation for the sample in Fig. 5(a) at $\lambda = 1580$ nm corresponding to a guided mode below the light line. It is evident that a strong scattering occurs at the termination of VPHW and is located at the center region, indicating that the scattering originates from the propagation of the guided mode. By probing the intensity of light scattering along the termination interface as red dashed line in Fig. 5(b), the profile of mode intensity is obtained and plotted in Fig. 5(c). It can be observed that the mode intensity is strongly localized at the center and exhibits rapid decay toward the surrounding regions. We also plotted mode intensity obtained from simulation as black line in Fig. 5(c), where measured data overlaps with simulated results, indicating consistency between the measurement and simulation. It is known that the resolution in measurement is limited by diffraction limit of the objective lens and wavelength [54]. For consistent comparison, the simulated intensity profiles were convolved with a Gaussian beam (FWHM=1.2 μm), reflecting the diffraction-limited resolution defined by the 1.55 μm wavelength and an objective lens with N.A.=0.65. Figure 5(d) and (e) show the mode intensity profiles for the guided modes in VPHW with $N = 2,4,8,16$ at $\lambda = 1580$ nm, obtained from simulation and measurement. Both simulation and measurement results show that the mode becomes broader as $N$ increases, suggesting that the mode width of guided modes below the light line is tunable through adjustment of $N$. To evaluate the mode widths, we calculated the FWHM of mode intensity profiles and plotted their dependance on $N$ in Fig. 5(f). Before calculating the FWHM of the measured intensity profiles, we first removed the background contribution caused by the reflection of the input light. To obtain the background intensity profiles, the input light was deliberately shifted to a position where no scattering occurs at the end of the VPhC heterostructure waveguides. From Fig. 5(f), it is observed that there is an approximately linear relationship between mode widths and $N$, and the measurement data fits well with simulation results. Our results provide the first experimental demonstration of tunable mode widths for guided modes below the light line in VPHW, offering new insights into the underlying physics of VPHWs and expanding their potential for low-loss, high-power photonic applications.

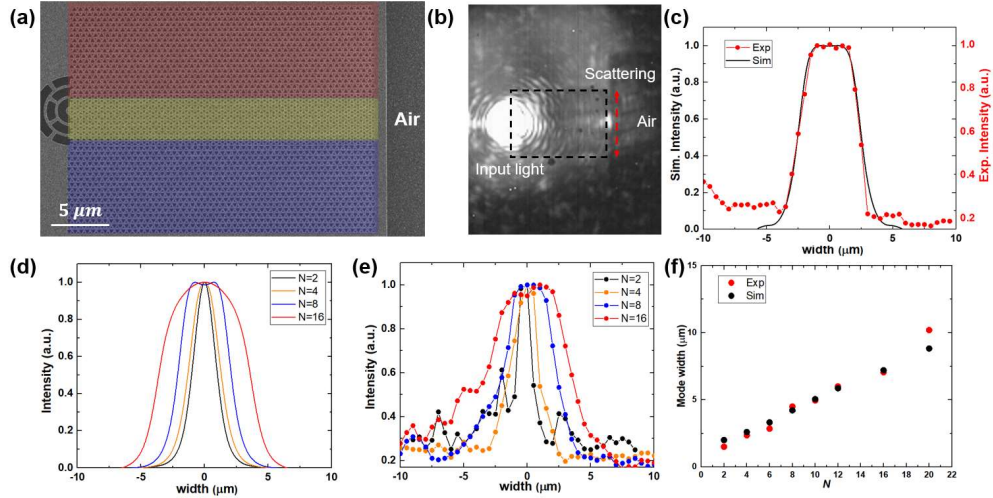

Fig. 5. (a) SEM image of fabricated VPhC heterostructure waveguide ($N = 10, d = 1.4a$) with termination of air slot. (b) Near-infrared microscope image of light scattering at the PhC-air interface for sample in (a). The sample position is indicated by a black dashed box. (c) Measured intensity profile for light scattering along red dashed line in (b). The simulation result is included as a comparison. Left and right vertical axes correspond to scales used for simulation and measured intensity profiles. (d) (e) Intensity profiles in simulation and measurement for guided modes in VPHW with $N = 2,4,8,16$ at $\lambda = 1580$ nm. (f) Relationship between mode widths for guide modes in VPHW and number of layers $N$ for simulation and measurement at $\lambda = 1580$ nm.

As discussed above, we measured the intensity profiles for guided modes with a linear dispersion below the light line in VPHWs. To demonstrate the wide-mode-area slow-light modes, utilizing the same method, we measured intensity profiles of slow-light guided modes below the light line in VPHWs. We fabricated VPHWs terminated by air slot for (a) $N = 4, d = 1.2a$, (b) $N = 6, d = 1.15a$, (c) $N = 8, d = 1.1a$, (d) $N = 10, d = 1.1a$, which support slow-light modes as group indices measurement (See Supplement 1). To measure the mode widths for slow-light modes with the same group index in VPHW for different $N$, we selected four modes with $n_g = 20$, which are $\lambda = 1535$ nm, 1532 nm, 1545 nm, 1545 nm for $N = 4,6,8,10$. Figure 6 depicts a near-infrared microscope image for light scattering in VPHW with $N = 8$ at $\lambda = 1545$ nm, which shows a scattering in the interface between VPHW and air slot. Similarly, we measured the mode intensity profiles along the VPHW-air interface indicated by red dashed line in Fig. 6(a). The measured and simulated mode intensity profiles for slow-light modes with $n_g = 20$ for VPHW with $N = 4,6,8,10$ are plotted in Fig. 6(b) and (c). It can be observed from both the simulated and measured profiles that the mode width increases with larger $N$, indicating that the mode width of slow-light modes in VPHWs is tunable by adjusting the number of PhC B layers. Notably, for $N = 10$, the slow-light mode exhibits a much broader mode width compared to that in interface-based waveguides, demonstrating the capability of VPHWs to support wide-mode-area slow-light modes. Figure 6(d) shows the mode widths evaluated by the FWHM of intensity profiles as a function of $N$ and indicates close agreement between simulation and experimental results.

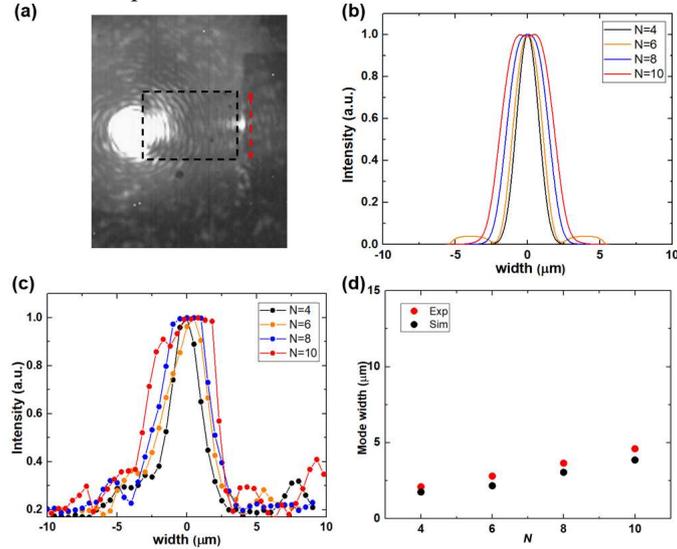

Fig. 6. (a) Near-infrared microscope image of light scattering for VPHW with $N = 8, d = 1.1a$ at $\lambda = 1545nm$. (b) (c) Intensity profiles in simulation and measurement for slow-light guided modes with $n_g = 20$ in VPHW with $N = 4,6,8,10$. (d) Relationship between mode widths for guide modes in VPHW and number of layers $N$ for simulation and measurement results in (b) and (c).

## 3. Conclusion

In conclusion, we have experimentally realized that wide-mode-area slow-light modes can be supported in valley photonic crystal (VPhC) heterostructure waveguides. Specifically, we fabricated VPhC heterostructure waveguides in a silicon slab with varying numbers of inserted gapless photonic graphene layers and different distances between photonic crystal unit cells near the domain walls. We found that the group index of guided modes increases as the distance between unit cells at the domain boundaries is reduced, enabling the realization of slow-light modes in structures with different numbers of photonic graphene layers. Optical microscopy revealed that the radiation observed from leaky guided modes becomes broader as more photonic graphene layers are inserted. To probe guided modes below the light line, we terminated the VPhC heterostructure waveguides with air slots, allowing scattering at the waveguide ends to be utilized for intensity profile measurements. The measured mode widths of both fast-light and slow-light guided modes were shown to be tunable by adjusting the number of photonic graphene layers. These findings open a promising route toward the development of photonic integrated circuits that incorporate on-chip slow-light devices capable of handling large optical power.

**Funding.** This work is supported by JST CREST(JPMJCR19T1), KAKENHI (22H00298, 22H01994) and Asahi Glass Foundation.

**Acknowledgments.** The authors thank Dr. Sangmin Ji and Mr. Nao Harada in The University of Tokyo and Dr. Siyuan Gao in Keio University for help in experiments.

**Disclosures.** The authors declare no conflicts of interest.

**Data availability.** Data underlying the results presented in this paper are not publicly available at this time but may be obtained from the authors upon reasonable request.

**Supplemental document.** See supplement 1 for supporting content.

## References

1. L. Lu, J. D. Joannopoulos, and M. Soljačić, "Topological photonics," Nat. Photonics **8**, 821-829 (2014).
2. T. Ozawa, H. M. Price, A. Amo, *et al.*, "Topological photonics," Rev. Mod. Phys. **91**, 015006 (2019).
3. A. B. Khanikaev and A. Alù, "Topological photonics: robustness and beyond," Nat. Commun. **15**, 931 (2024).
4. M. Jalali Mehrabad, S. Mittal, and M. Hafezi, "Topological photonics: Fundamental concepts, recent developments, and future directions," Physical Review A **108**, 040101 (2023).
5. H. Price, Y. Chong, A. Khanikaev, *et al.*, "Roadmap on topological photonics," J. Phys.: Photonics **4**, 032501 (2022).
6. Y. Wu, C. Li, X. Hu, *et al.*, "Applications of Topological Photonics in Integrated Photonic Devices," Adv. Opt. Mater. **5**, 1700357 (2017).
7. B.-Y. Xie, H.-F. Wang, X.-Y. Zhu, *et al.*, "Photonics meets topology," Opt. Express **26**, 24531-24550 (2018).
8. M. S. Rider, S. J. Palmer, S. R. Pocock, *et al.*, "A perspective on topological nanophotonics: Current status and future challenges," J. Appl. Phys. **125**, 120901 (2019).
9. M. Kim, Z. Jacob, and J. Rho, "Recent advances in 2D, 3D and higher-order topological photonics," Light Sci. Appl. **9**, 130 (2020).
10. S. Iwamoto, Y. Ota, and Y. Arakawa, "Recent progress in topological waveguides and nanocavities in a semiconductor photonic crystal platform [Invited]," Opt. Mater. Express **11**, 319-337 (2021).
11. M. Hafezi, S. Mittal, J. Fan, *et al.*, "Imaging topological edge states in silicon photonics," Nat. Photonics **7**, 1001-1005 (2013).
12. S. Barik, A. Karasahin, C. Flower, *et al.*, "A topological quantum optics interface," Science **359**, 666-668 (2018).
13. S. Barik, H. Miyake, W. DeGottardi, *et al.*, "Two-dimensionally confined topological edge states in photonic crystals," New J. Phys. **18**, 113013 (2016).
14. F. D. M. Haldane and S. Raghu, "Possible Realization of Directional Optical Waveguides in Photonic Crystals with Broken Time-Reversal Symmetry," Phys. Rev. Lett. **100**, 013904 (2008).
15. S. Raghu and F. D. M. Haldane, "Analogs of quantum-Hall-effect edge states in photonic crystals," Physical Review A **78**, 033834 (2008).
16. X. Wu, Y. Meng, J. Tian, *et al.*, "Direct observation of valley-polarized topological edge states in designer surface plasmon crystals," Nat. Commun. **8**, 1304 (2017).
17. Z. Gao, Z. Yang, F. Gao, *et al.*, "Valley surface-wave photonic crystal and its bulk/edge transport," Physical Review B **96**, 201402 (2017).
18. Z. Wang, Y. Chong, J. D. Joannopoulos, *et al.*, "Observation of unidirectional backscattering-immune topological electromagnetic states," Nature **461**, 772-775 (2009).
19. X. Cheng, C. Jouvaud, X. Ni, *et al.*, "Robust reconfigurable electromagnetic pathways within a photonic topological insulator," Nat. Mater. **15**, 542-548 (2016).
20. L.-H. Wu and X. Hu, "Scheme for Achieving a Topological Photonic Crystal by Using Dielectric Material," Phys. Rev. Lett. **114**, 223901 (2015).
21. F. Gao, H. Xue, Z. Yang, *et al.*, "Topologically protected refraction of robust kink states in valley photonic crystals," Nat. Phys. **14**, 140-144 (2018).
22. J.-W. Dong, X.-D. Chen, H. Zhu, *et al.*, "Valley photonic crystals for control of spin and topology," Nat. Mater. **16**, 298-302 (2017).
23. C. A. Rosiek, G. Arregui, A. Vladimirova, *et al.*, "Observation of strong backscattering in valley-Hall photonic topological interface modes," Nat. Photonics **17**, 386-392 (2023).

24. X.-D. Chen, F.-L. Shi, H. Liu, *et al.*, "Tunable Electromagnetic Flow Control in Valley Photonic Crystal Waveguides," Phys. Rev. Appl. **10**, 044002 (2018).
25. T. Yamaguchi, Y. Ota, R. Katsumi, *et al.*, "GaAs valley photonic crystal waveguide with light-emitting InAs quantum dots," Appl. Phys. Express **12**, 062005 (2019).
26. H. Yoshimi, T. Yamaguchi, Y. Ota, *et al.*, "Slow light waveguides in topological valley photonic crystals," Opt. Lett. **45**, 2648-2651 (2020).
27. H. Yoshimi, T. Yamaguchi, R. Katsumi, *et al.*, "Experimental demonstration of topological slow light waveguides in valley photonic crystals," Opt. Express **29**, 13441-13450 (2021).
28. P. Zhang, J. Zhang, L. Gu, *et al.*, "Compact on-chip power splitter based on topological photonic crystal," Opt. Mater. Express **14**, 1390-1397 (2024).
29. X.-T. He, C.-H. Guo, G.-J. Tang, *et al.*, "Topological Polarization Beam Splitter in Dual-Polarization All-Dielectric Valley Photonic Crystals," Phys. Rev. Appl. **18**, 044080 (2022).
30. G. Guo, H. Wang, Q. Wang, *et al.*, "Topologically protected power divider and wavelength division multiplexer based on valley photonic crystals," Opt. Express **33**, 12240-12252 (2025).
31. L. He, H. Zhang, W. Zhang, *et al.*, "Topologically protected vector edge states and polarization beam splitter by all-dielectric valley photonic crystal slabs," New J. Phys. **23**, 093026 (2021).
32. Y. Gong, S. Wong, A. J. Bennett, *et al.*, "Topological Insulator Laser Using Valley-Hall Photonic Crystals," ACS Photonics **7**, 2089-2097 (2020).
33. Y. Zeng, U. Chattopadhyay, B. Zhu, *et al.*, "Electrically pumped topological laser with valley edge modes," Nature **578**, 246-250 (2020).
34. X. Liu, L. Zhao, D. Zhang, *et al.*, "Topological cavity laser with valley edge states," Opt. Express **30**, 4965-4977 (2022).
35. M. Wang, R.-Y. Zhang, L. Zhang, *et al.*, "Topological One-Way Large-Area Waveguide States in Magnetic Photonic Crystals," Phys. Rev. Lett. **126**, 067401 (2021).
36. Q. Chen, L. Zhang, F. Chen, *et al.*, "Photonic Topological Valley-Locked Waveguides," ACS Photonics **8**, 1400-1406 (2021).
37. S. Yan, J. Yang, S. Shi, *et al.*, "Transport of a topologically protected photonic waveguide on-chip," Photonics Res. **11**, 1021-1028 (2023).
38. Q. Zhang, X. Xing, D. Zou, *et al.*, "Robust topological valley-locked waveguide transport in photonic heterostructures," Results Phys. **54**, 107066 (2023).
39. P.-Y. Guo, W. Li, J. Hu, *et al.*, "Dual-band topological large-area waveguide transport in photonic heterostructures," Physical Review B **110**, 035115 (2024).
40. W.-Y. Wang, H. Ren, Z.-H. Xu, *et al.*, "Integrated terahertz topological valley-locked power divider with arbitrary power ratios," Opt. Lett. **49**, 5579-5582 (2024).
41. W. Li, Q. Chen, Y. Sun, *et al.*, "Topologically Enabled On-Chip THz Taper-Free Waveguides," Adv. Opt. Mater. **11**, 2300764 (2023).
42. H. Shao, Y. Wang, G. Yang, *et al.*, "Topological transport in heterostructure of valley photonic crystals," Opt. Express **31**, 32393-32403 (2023).
43. Q. Zhang, X. Xing, D. Zou, *et al.*, "Investigation of unidirectional coupling of dipole emitters in valley photonic heterostructure waveguides," Opt. Express **32**, 415-424 (2024).
44. Z. Guan, K. Wen, C. Xie, *et al.*, "On-chip mode (de)multiplexer utilizing sandwich valley-topological edge waveguides," APL Photonics **10**, 036111 (2025).
45. P. Hu, L. Sun, C. Chen, *et al.*, "Arbitrary control of the flow of light using pseudomagnetic fields in photonic crystals at telecommunication wavelengths," (2025).
46. T. Baba, "Slow light in photonic crystals," Nat. Photonics **2**, 465-473 (2008).
47. H. Zhou, T. Gu, J. F. McMillan, *et al.*, "Enhanced four-wave mixing in graphene-silicon slow-light photonic crystal waveguides," Appl. Phys. Lett. **105**, 091111 (2014).
48. Y. A. Vlasov, M. O'Boyle, H. F. Hamann, *et al.*, "Active control of slow light on a chip with photonic crystal waveguides," Nature **438**, 65-69 (2005).
49. J. Liang, L.-Y. Ren, M.-J. Yun, *et al.*, "Wideband ultraflat slow light with large group index in a W1 photonic crystal waveguide," J. Appl. Phys. **110**, 063103 (2011).
50. Y. Hinakura, H. Arai, and T. Baba, "64 Gbps Si photonic crystal slow light modulator by electro-optic phase matching," Opt. Express **27**, 14321-14327 (2019).
51. C. Zhang, Y. Ota, and S. Iwamoto, "Wide-mode-area slow light waveguides in valley photonic crystal heterostructures," Opt. Mater. Express **14**, 1756-1766 (2024).
52. A. Faraon, I. Fushman, D. Englund, *et al.*, "Dipole induced transparency in waveguide coupled photonic crystal cavities," Opt. Express **16**, 12154-12162 (2008).
53. M. Notomi, K. Yamada, A. Shinya, *et al.*, "Extremely Large Group-Velocity Dispersion of Line-Defect Waveguides in Photonic Crystal Slabs," Phys. Rev. Lett. **87**, 253902 (2001).
54. I. H. A. Knottnerus, S. Pyatchenkov, O. Onishchenko, *et al.*, "Microscope objective for imaging atomic strontium with 0.63 micrometer resolution," Opt. Express **28**, 11106-11116 (2020).